\documentclass[iop,apj]{emulateapj}
\slugcomment{{\sc Accepted to ApJ:} November 11, 2012}

\usepackage{fontenc}
\usepackage{amssymb}
\usepackage{amsmath}
\usepackage{graphicx}
\usepackage{xspace}
\usepackage{microtype}
\usepackage{url}
\usepackage{paralist}
\usepackage{txfonts}
\usepackage{natbib}

\begin{document}

\title{The Be/X-ray binary Swift\,J1626.6$-$5156 as a
  variable cyclotron line source}

\author{Megan E.\ DeCesar\altaffilmark{1,2}}
\email{decesar@astro.umd.edu}

\author{Patricia T.\ Boyd\altaffilmark{3}}

\author{Katja Pottschmidt\altaffilmark{4,2}}

\author{J\"orn Wilms\altaffilmark{5}}

\author{Slawomir Suchy\altaffilmark{6,7}}

\author{M.\ Coleman Miller\altaffilmark{1,8}}

\altaffiltext{1}{Department of Astronomy, University of Maryland,
  College Park, MD 20742, USA}
  
\altaffiltext{2}{Center for Research and Exploration in Space Science
  and Technology, NASA-GSFC, Greenbelt, MD 20771, USA}

\altaffiltext{3}{NASA Goddard Space Flight Center, Code 661,
  Greenbelt, MD 20771, USA}

\altaffiltext{4}{Center for Space Science and Technology, University
  of Maryland, Baltimore County, 1000 Hilltop Circle, Baltimore, MD
  21250, USA}
  
\altaffiltext{5}{Dr.\ Karl Remeis-Sternwarte \& ECAP, University of
  Erlangen-Nuremberg, Sternwartstr.\ 7, 96049 Bamberg, Germany}
  
\altaffiltext{6}{Center for Astrophysics and Space Sciences,
  University of California, San Diego, 9500 Gilman Dr., La Jolla, CA
  92093-0424, USA}
 
  \altaffiltext{7}{Institut f\"ur Astronomie und Astrophysik,
    Abt.\ Astronomie, University of T\"ubingen, Sand 1, 72076
    T\"ubingen, Germany}

\altaffiltext{8}{Maryland Astronomy Center for Theory and Computation,
  University of Maryland, College Park, MD 20742, USA}

\begin{abstract}

  \noindent Swift\,J1626.6$-$5156 is a Be/X-ray binary that was in
  outburst from December 2005 until November 2008. We have examined
  \textsl{RXTE}/PCA and HEXTE spectra of three long observations of
  this source taken early in its outburst, when the PCA 2--20\,keV
  count rate was $>$70\,counts\,$\mathrm{s}^{-1}\,\mathrm{PCU}^{-1}$,
  as well as several combined observations from different stages of
  the outburst. The spectra are best fit with an absorbed cutoff power
  law with a $\sim 6.4$\,keV iron emission line and a Gaussian optical
  depth absorption line at $\sim$10\,keV. We present strong evidence
  that this absorption-like feature is a cyclotron resonance
  scattering feature, making Swift\,J1626.6$-$5156 a new candidate
  cyclotron line source. The redshifted energy of $\sim$10\,keV
  implies a magnetic field strength of $\sim 8.6(1+z) \times
  10^{11}$\,G in the region of the accretion column close to the
  magnetic poles where the cyclotron line is produced. Analysis of
  phase averaged spectra spanning the duration of the outburst
  suggests a possible positive correlation between the fundamental
  cyclotron energy and source luminosity. Phase resolved spectroscopy
  from a long observation reveals a variable cyclotron line energy,
  with phase dependence similar to a variety of other pulsars, as well
  as the first harmonic of the fundamental cyclotron line.
\end{abstract}


\section{Introduction}

\begin{figure}
\includegraphics[width=\columnwidth]{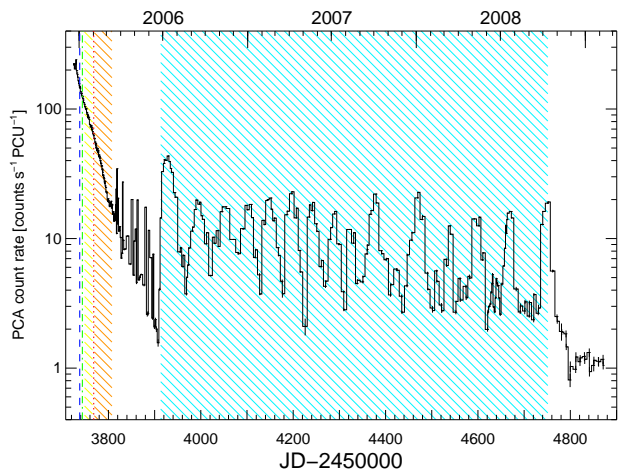}
\caption{PCA light curve of Swift\,J1626.6$-$5156, taken from
  HEASARC's \textsl{RXTE} mission-long data archive
  (ftp://legacy.gsfc.nasa.gov/xte/data/archive/MissionLongData). It
  shows the background subtracted 2--20\,keV count rate as recorded in
  \texttt{standard2} mode (16\,s time resolution), averaged over
  each observation and normalized to a single PCU. The three vertical
  lines indicate when the longest observations were taken in January
  2006.  LO1 is the blue dashed line, LO2 the green dot-dashed line,
  and LO3 the red dotted line. The shaded regions represent the Data
  Sections, with DS1 in yellow, DS2 in orange, and DS3 in
  teal. \label{fig1}}
\end{figure}

The high mass X-ray binary (HMXB) \objectname{Swift\,J1626.6$-$5156}
was discovered in outburst by the Swift Burst Alert Telescope (BAT) on
18 December 2005 \citep{2005krimm}. It was soon recognized to be an
X-ray pulsar with a $\sim$15\,s spin period \citep{2005palmer} and its
companion was classified as a B0Ve star at a distance of $\sim$10\,kpc
\citep{2006negueruela,2011reig}. Shortly after the discovery, the
Rossi X-ray Timing Explorer (\textsl{RXTE}) began monitoring the
source and continued to do so for nearly five years. The
categorization of Swift\,J1626.6$-$5156 as a Be/X-ray (BeX) binary
places it in a group that comprises most of the HMXB population, so we
will describe it in terms associated with that group of objects. Its
2--20\,keV PCA light curve is shown in Figure~\ref{fig1}, spanning the
beginning of \textsl{RXTE}'s monitoring in 2006 January through 2009
January, when the source had returned to quiescence.

\begin{deluxetable*}{cccccc}
\tablewidth{0pt}
\tablecaption{Observation Log \label{tab1}}
\tablecolumns{6} 
\tablehead{ \colhead{Name} & \colhead{Observation ID} & \colhead{Date}
  & \colhead{$\mathrm{JD} - 2450000$} & \colhead{PCA
    $t_{\mathrm{exp}}$} & \colhead{HEXTE $t_{\mathrm{exp}}$ [ks]} 
} \startdata
LO1 & 91081-13-01 & 2006 Jan 1 & 3736.05 & 26.4 & 8.4 \\
LO2 & 91081-13-02 & 2006 Jan 6 & 3741.15 & 23.6 & 7.4 \\
DS1 & 91082-01-01 -- 91082-01-27 & 2006 Jan 12--30 & 3747.73--3765.10 & 23.3 & 7.1 \\
LO3 & 91081-13-04 & 2006 Jan 31 & 3766.59 & 27.6 & 7.7 \\
DS2 & 91082-01-32 -- 91082-01-93 & 2006 Feb 1--Mar 13 &
3767.64--3807.52 & 51.2 & -- \\
DS3 & 92412-01-27 -- 93402-01-48 & 2006 Jun 26--2008 Oct 10 &
3912.21--4750.82 & 78.8 & -- 
\enddata
\end{deluxetable*}

Swift\,J1626.6$-$5156 was discovered during a Type~II outburst, which
is thought to occur when the Be star's circumstellar disk expands and
temporarily engulfs the neutron star, leading to enhanced accretion
and therefore a large increase in X-ray emission \citep[][and
references therein]{2000coe}. \citet{2008reig} describe the
exponential decay of the outburst and the flaring behavior that
occurred in the weeks immediately following the discovery. Additional
flaring occurred following JD 2453800 (2006 March 5), after which the
source transitioned into an epoch of long-term, quasi-periodic X-ray
oscillations \citep{2008reig}. A Lomb-Scargle periodogram
\citep{1976lomb,1982scargle} of the light curve first revealed an
oscillation period from $\sim$JD\,2454000--JD\,2454350 (2006
September--2007 September) of $\sim$47\,days and from
$\sim$JD\,2454350--2454790 (2007 September--2008 November) of
$\sim$3/2 of this, or $\sim$72.5\,days \citep{2009decesar}. Recent
work by \citet{2010baykal} found the long-term variability timescale
to range from $\sim$45 to 95\,days. Additionally, from pulse timing
analysis, they find the binary orbit to be nearly circular
(eccentricity 0.08) with a period of 132.9\,days, meaning the
oscillations occur on timescales of $\sim$1/2 and $\sim$2/3 of the
orbital period. The oscillations therefore cannot be explained by
bursts of accretion during periastron passages. On JD\,2454791 (2008
November 11) the oscillatory behavior ceased and the 2--20\,keV PCA
count rate from Swift\,J1626.6$-$5156 dropped to
$\sim$$1\,\mathrm{count}\,\mathrm{s}^{-1}\,\mathrm{PCU}^{-1}$.

Many BeX and other HMXB systems harbor neutron stars with very strong
magnetic fields, typical strengths being $\sim 10^{11-13}$\,G, and
some of these systems exhibit cyclotron resonance scattering features
(CRSFs or cyclotron lines). Observing cyclotron lines in the spectra
of these systems is the only direct way to measure the magnetic field
strengths of neutron stars. At the time of writing there are sixteen
neutron star X-ray binaries with confirmed CRSFs \citep[][and
references therein]{2004heindl,2012caballero,2012pottschmidt}, eight
of which are transients (Swift\,J1626.6$-$5156 being the ninth). Of
the transients four systems are BeX binaries (Swift\,J1626.6$-$5156
being the fifth) and four are OeX binaries. The persistent sources are
HMXBs with an O- or B-star accompanying the neutron star (with
exception of Her X-1 and 4U\,1626$-$67, where the companion is of
lower mass). The fundamental cyclotron line energy is related to the
magnetic field strength of the neutron star $E_\mathrm{cyc} \sim
11.6\,B_{12} (1+z)^{-1}$\,keV where $B_{12} = B /10^{12}\,\mathrm{G}$
and $z$ is the surface gravitational redshift. Assuming a reasonable
value of $z=0.3$, corresponding to a typical neutron star mass of $1.4
M_\odot$, allows the direct calculation of $B$ in the region local to
the magnetic pole where the cyclotron lines are produced.

In this Paper we discuss the discovery of a CRSF in the spectrum of
Swift\,J1626.6$-$5156. In \S\ref{sec:observations} we describe our
data reduction and analysis techniques. \S\ref{sec:spectra} presents
results from modeling the broad band X-ray spectra, including CRSFs
and other spectral properties of the source. We summarize our findings
and compare Swift\,J1626.6$-$5156 with other cyclotron line sources in
\S\ref{sec:discussion} and conclude in \S\ref{sec:conclusion}.

\section{Observations and Data Analysis}\label{sec:observations}

For this analysis, we considered data from the Proportional Counter
Array \citep[PCA;][]{2006jahoda} and the High Energy X-ray Timing
Experiment \citep[HEXTE;][]{1998rothschild} onboard \textsl{RXTE} that
were taken between December 2005 and April 2009. From HEXTE, we used
spectra from only the B cluster, since the A cluster had stopped
rocking at the time of these observations\footnote{See
  \url{http://heasarc.gsfc.nasa.gov/docs/xte/whatsnew/big.html} for
  more detailed information.}, i.e., no background measurements for
cluster A were available anymore. We extracted spectra in
\texttt{standard2} mode from the top layer of each PCA Proportional
Counter Unit (PCU) separately. We then combined all spectra from a
given observation into one, following the recipe in the \textsl{RXTE}
Cookbook\footnote{The \textsl{RXTE} Cookbook can be found at\\
  \url{http://heasarc.gsfc.nasa.gov/docs/xte/recipes/cook\_book.html}.}
and excluding data taken within 10\,min of the South Atlantic Anomaly.
For a more detailed description of the data reduction and screening
procedure see \citet{2006wilms}. We considered the PCA energy range
4--22\,keV and the HEXTE range 18--60\,keV when modeling the phase
averaged spectra.

Three long observations with PCA exposure time $>$20\,ks are available
in the \textsl{RXTE} archive, along with hundreds of shorter
monitoring observations. We combined the individual pointings of each
observation as described in the \textsl{RXTE} Cookbook. The three
``Long Observations'', which we refer to as LO1, LO2, and LO3, were
each analyzed separately, while the data between and beyond these
observations were combined into ``Data Sections'', referred to as DS1,
DS2, and DS3. Data prior to LO1 were excluded in order to avoid
spectral contamination from flaring events \citep[for a description of
the flares, see][]{2008reig}. For DS3 we combined only observations from
the oscillating phase that had an average PCA 2--20\,keV count rate of
10\,$\mathrm{counts}\,\mathrm{s}^{-1}\,\mathrm{PCU}^{-1}$ or higher.
The observation identification numbers and other information are given
in Table~\ref{tab1}. The locations in time of LO1--3 are shown by the
colored lines and of DS1--3 by the shaded regions in
Figure~\ref{fig1}. The HEXTE data have low count rates, so we rebinned
the counts at high energies using the \texttt{FTOOL}
\texttt{grppha}\footnote{See
  \url{http://heasarc.nasa.gov/ftools/ftools\_menu.html} for more
  detailed information.}. For LO1 and LO2 we rebinned by three energy
channels above 50\,keV and for LO3 by four channels above 48\,keV. We
did not rebin any counts for DS1. The HEXTE data of DS2 and DS3 were
so noisy we decided not to use them as they might skew our spectral
fits. We used \texttt{xspec12} models \citep{xspec,xspec2} for all
spectral fitting\footnote{The manual can be found at \\
  \url{http://heasarc.nasa.gov/xanadu/xspec/manual/manual.html}.}.

\section{Spectral Analysis}\label{sec:spectra}

\subsection{Spectral Model}\label{sec:model}

We modeled the combined PCA and HEXTE (where present) phase averaged
spectra of each LO and DS with a power law modified by a
  smooth exponential rollover (\texttt{cutoffpl} in \texttt{xspec})
including photoelectric absorption (\texttt{phabs} in \texttt{xspec})
and a Gaussian 6.4\,keV Fe K$\alpha$ emission line (\texttt{gauss} in
\texttt{xspec}):

\begin{eqnarray}
\nonumber I_{\mathrm{cont+Fe}}(E) & = & e^{-N_{\mathrm{H}} \sigma_{\mathrm{bf}}(E)}  \left \{  A_{\mathrm{cut}} E^{-\Gamma} e^{-\frac{E}{E_{\mathrm{fold}}}} \right. \\
                     & & + \left. \frac{ A_{\mathrm{Fe}}}{\sigma_{\mathrm{Fe}}\sqrt{2\pi} }
          e^{-\frac{1}{2}[(E-E_{\mathrm{Fe}})^2 /
          \sigma_{\mathrm{Fe}}^2]} \right \}
\label{eq1}
\end{eqnarray}
where $N_{\mathrm{H}}$ and $\sigma_{\mathrm{bf}}$ are the absorption
model components and are respectively the hydrogen column density per
H atom for material of cosmic abundance
($\mathrm{atoms}\,\mathrm{cm}^{-2}$) and the bound-free photoelectric
absorption cross section ($\mathrm{cm}^2$). We used the abundances of
\citet{1989angr} and the cross sections of \citet{1992bcmc}.

The first term in Eq.~\ref{eq1} is the cutoff power law, where
$A_\mathrm{cut}$ is the power law normalization
($\mathrm{photons}\,\mathrm{keV}^{-1}\,\mathrm{cm}^{-2}\,\mathrm{s}^{-1}$
at 1\,keV), $\Gamma$ is the power law photon index, and
$E_{\mathrm{fold}}$ is the e-folding energy of the exponential rollover.
The second term is the Gaussian emission line, in which
$A_{\mathrm{Fe}}$ is its normalization
($\mathrm{photons}\,\mathrm{cm}^{-2}\,\mathrm{s}^{-1}$),
$E_\mathrm{Fe}$ its energy, and $\sigma_\mathrm{Fe}$ its width. This
model was applied concurrently to the PCA and HEXTE data, differing
only by a constant flux cross-calibration factor where the
  constant was fixed at 1 for the PCA and fit to $\sim$0.8 for HEXTE.
For each spectral fit we also fit the background strength adjusting it
on a level of $\sim$0.1\% using the command \texttt{recornorm} in
\texttt{xspec}.

\begin{figure}
\includegraphics[width=\columnwidth]{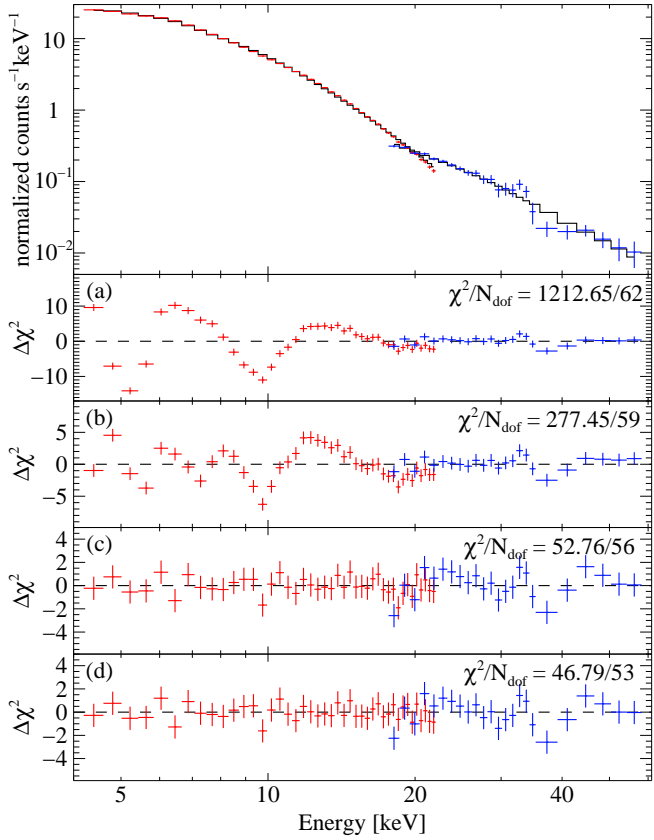}
\caption{The $\sim$4--60\,keV spectrum and best-fit model for
  observation LO3 are shown in the main panel. PCA data are shown in
  red, HEXTE data in blue. \textit{a)} The $\Delta \chi^2$ residuals
  left from fitting the spectrum with a cutoff power law. The spectrum
  peaks near $6.4\,$keV and dips near $10\,$keV. \textit{b)} Residuals
  from adding a Gaussian emission line at $\sim 6.4\,$keV to the
  cutoff power law, completing our continuum plus Fe line fit from
  Equation~\ref{eq1}. The depression at $10$\,keV remains. \textit{c)}
  Including a Gaussian optical depth absorption line near $10$\,keV
  removes the absorption-like feature and greatly improves the model,
  giving our best fit (Equation~\ref{eq2}). There may be a second
  feature near $18$\,keV. \textit{d)} A second Gaussian optical depth
  absorption feature included near $18$\,keV marginally improves the
  fit. The remaining residual near $\sim 36$\,keV is most likely caused by 
  an instrumental effect (\S\ref{sec:iodine}).\label{fig2}}
\end{figure}

\begin{figure}
\includegraphics[width=\columnwidth]{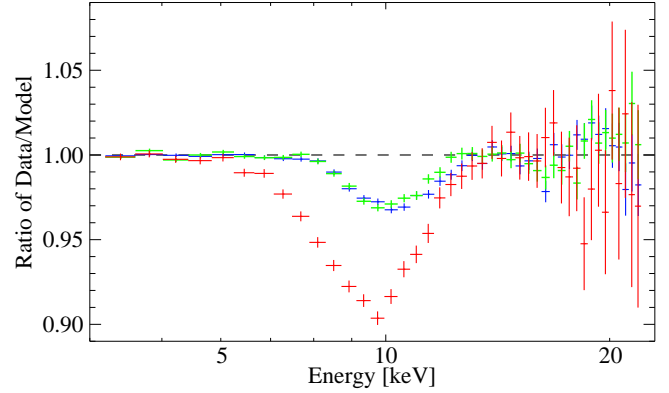}
\caption{Removing the Gaussian optical depth absorption component in
  our spectral model reveals the profile of the fundamental cyclotron
  line at $\sim$10\,keV in LO1 (blue), LO2 (green), and LO3 (red). A
  decrease of the central energy between LO1--LO2 (close in time) and
  LO3 is apparent, as well as an increase in line depth. \label{fig3}}
\end{figure}

Residuals for a continuum fit using the model from Eq.~\ref{eq1}
without and with the iron emission line for LO3 are shown in
Figs~\ref{fig2}a and~\ref{fig2}b, respectively. A depression in the
residuals at $\sim$10\,keV is apparent. In order to account for this
feature, we modified the continuum plus Fe line model with a
multiplicative absorption-like cyclotron line with a
Gaussian optical depth profile (\texttt{gabs} in \texttt{xspec}):
\begin{equation}
I(E) = I_{\mathrm{cont+Fe}} (E) \exp \left \{ 
\frac{- \tau_{\mathrm{cyc}}}{\sigma_{\mathrm{cyc}} \sqrt{2\pi}} \exp 
 \left [ - \frac{(E-E_\mathrm{cyc})^2}{2 \sigma_{\mathrm{cyc}}^2}
   \right ] \right \}
\label{eq2}
\end{equation}
Here $\tau_{\mathrm{cyc}}$, $\sigma_{\mathrm{cyc}}$, and $E_\mathrm{cyc}$
are respectively the optical depth, line width, and line energy of the
cyclotron line. The residuals of a fit to LO3 using the full model
$I(E)$ of Eq.~\ref{eq2}, i.e., including the cyclotron line, are
displayed in Figure~\ref{fig2}c.

In the following sections we describe the modeling and interpretation
of this line as a CRSF including several tests that led to the
adoption of the model described above as the best-fit model for the
phase averaged spectra
(\texttt{constant}$\times$\texttt{phabs}$\times$(\texttt{cutoffpl} +
\texttt{gauss})$\times$\texttt{gabs} in \texttt{xspec}).

\begin{deluxetable*}{ccccccccccccccc}
\tablewidth{0pt}
\tabletypesize{\scriptsize}
\scriptsize
\tablecaption{Spectral parameters for phase averaged spectra \label{tab2} }
\tablecolumns{14} \tablehead{
\colhead{Name} & \colhead{$N_{\mathrm{H}}$} & \colhead{$\Gamma$} & \colhead{$A_\mathrm{cut}$} &
\colhead{$E_{\mathrm{fold}}$} & \colhead{$E_{\mathrm{Fe}}$} &
\colhead{$\sigma_{\mathrm{Fe}}$} & \colhead{$A_{\mathrm{Fe}}$} & 
 \colhead{$E_\mathrm{cyc}$} &
\colhead{$\sigma_{\mathrm{cyc}}$} &
\colhead{$\tau_{\mathrm{cyc}}$} &  \colhead{$F$} &
\colhead{$\chi^2/N_{\mathrm{dof}}$} \\
\colhead{} & \colhead{[$10^{22}$]$^b$} & \colhead{} &
\colhead{[$10^{-1}$]$^c$} & \colhead{[keV]} & \colhead{[keV]} & \colhead{[keV]} &
\colhead{[$10^{-3}$]$^d$}  & \colhead{[keV]} &  \colhead{[keV]}
& \colhead{} &   \colhead{[$10^{-10}\, \mathrm{cgs}$]} & \colhead{}
} \startdata
LO1 & $4.8^{+0.6}_{-0.6}$ & $1.11^{+0.04}_{-0.05}$ & $2.8^{+0.2}_{-0.3}$ &
$10.46^{+0.33}_{-0.34}$ & $6.33^{+0.03}_{-0.04}$ &
$0.49^{+0.05}_{-0.05}$ & $3.3^{+0.5}_{-0.4}$  &
$10.23^{+0.09}_{-0.09}$ & $0.99^{+0.17}_{-0.17}$ &
$0.097^{+0.025}_{-0.021}$ &  $19.10^{+0.02}_{-0.10}$
& $81.9/68$ \\ [1mm]
LO2 & $4.5^{+0.8}_{-0.7}$ & $1.20^{+0.06}_{-0.05}$ & $2.5^{+0.3}_{-0.2}$ &
$11.50^{+0.52}_{-0.39}$ & $6.30^{+0.04}_{-0.04}$ &
$0.54^{+0.07}_{-0.05}$ & $3.0^{+0.4}_{-0.5}$ &
$10.01^{+0.09}_{-0.08}$ & $0.76^{+0.15}_{-0.16}$ &
$0.082^{+0.020}_{-0.017}$ &  $16.80^{+0.03}_{-0.20}$
& $78.9/68$ \\ [1mm]
DS1 & $4.8^{+0.8}_{-0.7}$ & $1.37^{+0.07}_{-0.06}$ & $2.4^{+0.3}_{-0.3}$ &
$12.72^{+0.79}_{-0.37}$ & $6.34^{+0.05}_{-0.04}$ &
$0.45^{+0.06}_{-0.08}$ & $1.7^{+0.3}_{-0.3}$  &
$9.74^{+0.07}_{-0.07}$ & $0.96^{+0.13}_{-0.12}$  &
$0.15^{+0.02}_{-0.02}$                &  $11.80^{+0.04}_{-0.20}$
& $94.1/76$ \\ [1mm]
LO3 & $3.4^{+0.7}_{-0.8}$ & $1.32^{+0.09}_{-0.09}$ & $1.7^{+0.2}_{-0.2}$ &
$10.98^{+0.97}_{-0.91}$ & $6.38^{+0.04}_{-0.04}$ &
$0.19^{+0.10}_{-0.18}$ & $0.9^{+0.2}_{-0.1}$  &
$9.71^{+0.07}_{-0.07}$ & $1.27^{+0.16}_{-0.15}$  &
$0.33^{+0.04}_{-0.06}$               &  $8.62^{+0.04}_{-0.19}$
& $52.8/56$ \\ [1mm]
DS2 & $1.4^{+0.8}_{-1.2}$ & $1.61^{+0.07}_{-0.08}$ & $1.0^{+0.1}_{-0.1}$ &
$19.07^{+1.83}_{-2.47}$ & $6.36^{+0.03}_{-0.06}$ &
$0.41^{+0.05}_{-0.06}$ & $0.8^{+0.2}_{-0.1}$  &
$10.00^{+0.08}_{-0.08}$ & $0.64^{+0.02}_{-0.02}$ &
$0.25^{+0.04}_{-0.04}$               &  $4.44^{+0.03}_{-0.15}$
& $59.4/33$ \\ [1mm]
DS3$^a$ & --              & $1.37^{+0.07}_{-0.07}$ &
$0.33^{+0.02}_{-0.02}$ & $13.20^{+1.67}_{-1.30}$ &
$6.48^{+0.02}_{-0.02}$ & $0.41^{+0.05}_{-0.06}$ &
$0.40^{+0.04}_{-0.04}$  & $9.44^{+0.07}_{-0.06}$ &
$0.64^{+0.11}_{-0.12}$ & $0.25^{+0.04}_{-0.04}$ &
$2.06^{+0.003}_{-0.04}$ & $52.4/32$ 
\enddata \tablecomments{The best-fit spectral parameters from each
  observation or group of observations. $N_{\mathrm{H}}$ is the column
  density, $\Gamma$ the power law photon index,
  $A_\mathrm{cut}$ the power law normalization,
  $E_\mathrm{fold}$ the folding energy of the cutoff power law,
  $E_\mathrm{Fe}$, $\sigma_\mathrm{Fe}$, and $A_\mathrm{Fe}$
  the iron emission line energy, width, and normalization,
  $E_\mathrm{cyc}$, $\sigma_\mathrm{cyc}$, and $\tau_\mathrm{cyc}$ the
  energy, width, and optical depth of the Gaussian optical depth
  absorption line profile used to fit the cyclotron line, $F$ the
  3--20\,keV source flux in units of
  $10^{-10}\,\mathrm{erg}\,\mathrm{cm}^{-2}\,\mathrm{s}^{-1}$, and
  $\chi^2/N_\mathrm{dof}$ the reduced $\chi^2$ (where $N_\mathrm{dof}$
  is the number of degrees of freedom), of each spectral fit. The
  changing $N_\mathrm{dof}$ results from differences in rebinning of
  the HEXTE data (see \S\ref{sec:observations}) for the LO1--LO3
  columns, and from our choice to not use HEXTE data for DS2--3. The
  error bars quoted are at the 90\% level. The 90\% flux errors were
  calculated in \texttt{xspec} using 1000 draws from a Gaussian
  distribution. $^a$DS3 requires an additional Gaussian emission
  component with energy $15.18^{+0.17}_{-0.17}$\,keV, $\sigma$ fixed
  at 0.30\,keV, and normalization $7.3^{+1.4}_{-1.4} \times
  10^{-5}\,\mathrm{photons}\,\mathrm{cm}^{-2}\,\mathrm{s}^{-1}$.
  $^b$Values of $N_\mathrm{H}$ are in units of
  $10^{22}\,\mathrm{atoms}\,\mathrm{cm}^{-2}$. $^c$Units are
  $10^{-1}\,\mathrm{photons}\,\mathrm{keV}^{-1}\,\mathrm{cm}^{-2}\,\mathrm{s}^{-1}$
  at 1\,keV. 
  $^d$Values are given in units of $10^{-3}\,\mathrm{photons}\,\mathrm{cm}^{-2}\,\mathrm{s}^{-1}$. }
\end{deluxetable*}

\subsection{Pulse phase averaged spectra}\label{sec:phaseaveraged}

\subsubsection{Best-fit results, 10\,keV CRSF detection}\label{sec:best-fit}

We used LO3, the longest observation of Swift\,J1626.6$-$5156, to
illustrate the results obtained with the best-fit model described in
the previous section. This observation was taken on 2006 January 31,
during the outburst decay (Figure~\ref{fig1}). The first PCA pointing
is $\sim$21\,ks long and is followed immediately by a shorter pointing
of $\sim$6\,ks. We combined them to form a single spectrum with PCA
integration time $\sim$27\,ks. The corresponding HEXTE observation has
$\sim$7.6\,ks of live time. The best-fit parameters for the LO3
spectrum using the model of Eq.~\ref{eq2}, along with the iron line
equivalent width and other characteristics, can be found in the fourth
row of Table~\ref{tab2}. The phase averaged spectrum of LO3 and
best-fit model are shown in the main panel of Figure~\ref{fig2}.
Inspecting the residuals obtained after fitting the continuum model
and iron line (Equation~\ref{eq1}) shown in Figure~\ref{fig2}b more
closely reveals the absorption-like feature near 10\,keV and/or an
emission-like feature near 14\,keV. Here we treated the feature as
absorption-like only (Equation~\ref{eq2}) and obtained a very good
fit. See \S\ref{sec:crab}--\ref{sec:modified} for testing possible
alternative descriptions. Fitting the feature with a Gaussian optical
depth absorption line flattened the residuals in
Figure~\ref{fig2}c, yielding $\chi^2_{\mathrm{red}} = 0.94$
  for 56 degrees of freedom. The common $\Delta \chi^2$ test available
  in \texttt{xspec} could not be applied in this case, as it is not
  applicable to multiplicative model components
  \citep{2012orlandini}. We instead used the more appropriate ratio of
  variances $F$-test \citep{2007press} to calculate the probability
  that including this absorption-like feature in the model improves
  the fit by chance. As discussed by \citet{2012orlandini}, this
  $F$-test is not available in \texttt{xspec}; throughout this paper,
  we use the IDL routine
  \texttt{mpftest}\footnote{\url{http://www.physics.wisc.edu/$\sim$craigm/idl/down/mpftest.pro}}
  to calculate the probability of chance improvement (PCI) from one
  model to another. If one model is not a significant improvement over
  another, then the PCI will be large.  The PCI of the model that
  includes the Gaussian-depth absorption feature (Figure~\ref{fig2}c)
  compared to that without (Figure~\ref{fig2}b) is $4.6 \times
  10^{-9}$.  The improvement to the fit due to use of the
  absorption-like model component is therefore very statistically
  significant. The best-fit parameters of this model are given in
Table~\ref{tab2}.



The CRSF is detected in the earlier two long outburst observations,
LO1 and LO2, as well. Figure~\ref{fig3} shows the line profiles for
LO1--LO3. We also detected a strong iron line in each of the spectra.
See Table~\ref{tab2} for the best-fit results. One question we may ask
is whether or not the cyclotron line persists throughout and beyond
the primary outburst decay. To answer this we combined sections of
data into single spectra to search for the line. The data selection is
described in \S2 and summarized in Table~\ref{tab1}. We found that
both the iron line and the CRSF were present throughout the active
phase of Swift\,J1626.6$-$5156, as the features are detected
  in each of the Data Sections DS1--DS3 as well as in the Long
  Observations.
Interestingly, we found that the spectrum from the oscillatory stage
of the light curve (DS3) is best fit by the addition of a Gaussian
emission line at $\sim$15\,keV, a component that was not necessary in
any of the spectra taken during the outburst decay (see
\S\ref{sec:modified}) but that is commonly used to fit HMXB spectra
\citep{2002coburn}. All fit parameters are given in Table~\ref{tab2}.
The values of $E_\mathrm{cyc}$ are suggestive of an evolution of the
cyclotron line energy with luminosity, which is discussed in
\S\ref{sec:luminosity}. Note that the parameters might be affected by
additional uncertainties due to the range of fluxes covered for
DS1--DS3 and missing HEXTE data for DS2--DS3, though.

``Bumps'' or ``wiggles'' in the spectra of accreting X-ray pulsars,
especially near 10\,keV, are common and have been discussed at some
length by \citet{2002coburn}. They find residuals in the data/model
ratio at the $\sim$0.8\% level in all the accreting systems they
examine. These residuals could easily be interpreted as cyclotron
features. There have been marginal detections of absorption-like
features, possibly cyclotron lines, near 10\,keV in other sources. One
example is the recent work on \objectname{XMMU\,J054134.7$-$682550} by
\citet{2009inam}. These authors see what may be an absorption-like
feature at 10\,keV, but the count rate is too low to claim a
significant detection. One must thus take good care to ensure that the
feature we see is truly a cyclotron line and not due to calibration
uncertainties, additional emission (e.g., from the Galactic
  ridge), or incorrect modeling of the continuum. We note that the
  residuals we see at $\sim$10\,keV are on the 3--10\% level
  (Figure~\ref{fig3}), i.e., comparatively strong. Nevertheless we
  address these potential sources of uncertainties in
  \S\ref{sec:crab}, \S\ref{sec:ridge}, and
  \S\ref{sec:cont}--\S\ref{sec:modified}, respectively. We also note
  that \citet{2011reig} state that they do not detect a cyclotron line
  in the \textsl{RXTE} outburst data. However, these authors include a
  9\,keV edge in their model apparently masking the CRSF. This model
  choice misses the typical CRSF properties of the observed residual
  that we present in this paper (e.g., its luminosity or pulse phase
  dependence). 

\subsubsection{Crab spectrum comparison}\label{sec:crab}

\begin{figure}
\includegraphics[width=\columnwidth]{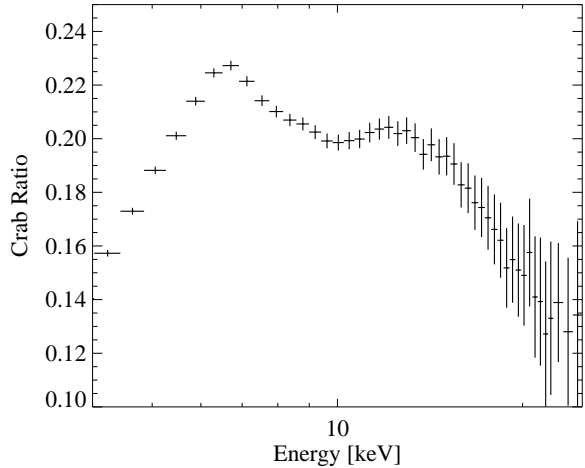}
\caption{Ratio of Swift\,J1626$-$5156 (observation LO3) to
    Crab counts according to the procedure described by
    \citet{1998orlandini}. The depression at $\sim$10\,keV is present
    in this model- and calibration-independent representation of the
    spectrum of Swift\,J1626$-$5156 as well. \label{fig4}}
\end{figure}

One test we can perform in order to validate the existence of the
cyclotron line in the phase averaged spectrum is to compare the
spectrum of Swift\,J1626.6$-$5156 with that of a non-cyclotron line
source like the Crab pulsar. To be as consistent as possible in our
comparison, we used Swift\,J1626.6$-$5156 and Crab data only from
PCU~2, which is the best calibrated of the five PCUs
\citep{2006jahoda} and which is most often turned on during
observations. The Crab data included both the pulsar and the nebula,
and were taken between November 2005 and March 2006, the same time
period during which the Swift\,J1626.6$-$5156 observations LO1--3 were
taken.

We first searched for an absorption-like feature near 10\,keV in the
Crab spectral residuals. The Crab has no CRSFs, so if such a feature
were seen in the Crab spectrum, we could assume that the
Swift\,J1626.6$-$5156 feature was instrumental in origin. We modeled
the Crab spectrum with an absorbed power law, freezing
$N_{\mathrm{H}}$ at $4 \times 10^{21}\,\mathrm{cm}^{-2}$, the value
determined for the Crab by \citet{2004weisskopf}. The power law
parameters were left free, and we found that $\Gamma = 2.095$ with a
normalization of
10.343\,$\mathrm{photons}\,\mathrm{keV}^{-1}\,\mathrm{cm}^{-2}\,\mathrm{s}^{-1}$
at 1\,keV gave the best fit. The data/model ratio is $\sim$0.99 near 10\,keV, which is consistent
with expected calibration uncertainties \citep{2006jahoda}. No
absorption-like feature is found at 10\,keV in the Crab spectrum (and
no rollover). We repeated the procedure with PCU~2 data from
Swift\,J1626.6$-$5156 during LO3, using a cutoff power law and fixing
$N_{\mathrm{H}}$ and $\Gamma$ to the values in Table~\ref{tab2}. In
this case the ratio near 10\,keV is $\sim$0.91, a much larger residual
than consistent with calibration errors.

\begin{deluxetable}{cccccc}
\tablewidth{0pt}
\tabletypesize{\scriptsize}
\scriptsize
\tablecaption{Comparison with alternative continuum models \label{tab:new} }
\tablecolumns{6} \tablehead{
\colhead{Continuum Model} & \colhead{$E_{\mathrm{Fe}}$} & \colhead{$E_{\mathrm{cyc}}$} & \colhead{$\chi^2_{\mathrm{red}}$} & \colhead{$N_{\mathrm{dof}}$} & \colhead{PCI} \\
\colhead{} & \colhead{[keV]} & \colhead{[keV]} & \colhead{} & \colhead{} & \colhead{}
} \startdata
\texttt{cutoffpl} & $6.38^{+0.04}_{-0.04}$ & $9.71^{+0.07}_{-0.07}$ & $0.94$ & $56$ & -- \\ [1mm]
\texttt{power}$\times$\texttt{highecut} & $6.38^{+0.04}_{-0.04}$ & $9.74^{+0.08}_{-0.07}$ & $0.96$ & $55$ & $0.47$ \\ [1mm]
\texttt{power}$\times$\texttt{fdcut} & $6.38^{+0.04}_{-0.04}$ & $9.79^{+0.11}_{-0.12}$ & $1.12$ & $55$ & $0.26$ \\ [1mm]
\texttt{npex} & $6.38^{+0.06}_{-0.06}$ & $9.83^{+0.28}_{-0.25}$ & $1.02$ & $54$ & $0.38$ 
\enddata
\tablecomments{Comparison between the \texttt{cutoffpl} continuum model with the \texttt{highecut}, \texttt{fdcut}, and \texttt{npex} models described in \S\ref{sec:cont}. All spectral modeling was performed on the LO3 dataset. Fits with the alternative continuum models yield values of $E_{\mathrm{Fe}}$ and $E_{\mathrm{cyc}}$, defined as in Table~\ref{tab2}, that are consistent with the line energies found using the \texttt{cutoffpl} model.  The probability of chance improvement (PCI) is described in \S\ref{sec:cont} and calculated with $\chi^2_{\mathrm{red}}$ and $N_{\mathrm{dof}}$ of the \texttt{cutoffpl} and each of the alternative models.  The PCI is large for each of the three alternative models, meaning that the different continuum models are statistically equivalent. For ease of comparison, we repeat the selected results for the \texttt{cutoffpl} continuum from Table~\ref{tab2}.}
\end{deluxetable}

We next took a model and calibration independent approach by comparing
the raw counts of Swift\,J1626.6$-$5156 with those of the Crab in each
spectral energy bin following the procedure of \citet{1998orlandini}.
We first subtracted the background counts from each raw spectrum,
making sure to account for the difference in exposure time between the
two sources. We then divided the Swift\,J1626.6$-$5156 count rates by
those of the Crab, resulting in the ratio shown in
Figure~\ref{fig4}. At $\lesssim$6\,keV the J1626/Crab ratio reflects
the steeper slope of the Crab and at $\gtrsim$14\,keV the missing Crab
rollover. Between these energies an enhancement at 6.4\,keV due to
the  iron line in Swift\,J1626.6$-$5156 and a clear depression at
$\sim$10\,keV, rather than a smooth transition between the iron line
and the power law tail, are apparent. The same kind of Crab
ratio is shown, e.g., for the $\sim$55\,keV CRSF of \objectname{Vela
X-1} in the upper panel of Figure~3 of \citet{1998orlandini}. This
plot shows similar cyclotron line and rollover features as the
J1626/Crab ratio (it differs at lower energies due to the overall
different energy range). Independently of the response function or
spectral model the J1626/Crab ratio therefore supports that the
absorption-like feature does not result from a calibration error.

\subsubsection{Galactic ridge emission}\label{sec:ridge}

We also investigated the possibility that cyclotron features
  in the spectrum arise from the Galactic ridge emission by analyzing
  data from when the source was quiescent. While Swift\,J1626.6$-$2126
  remained fairly bright and highly variable for about 1000\,days
  after the initial outburst, it then quickly faded to quiescence
  (Figure~\ref{fig1}) and stayed reliably low for the last two years of
  the \textsl{RXTE} mission. During this time the source continued to
  be monitored regularly by the PCA. We combined these late
  observations to obtain a spectrum with a total of 42\,ks exposure,
  and compared this with our outburst observations, in order to put
  limits on the contribution of the flux during the outburst
  observations from the underlying Galactic ridge emission. In the
  10\,keV region surrounding the fundamental cyclotron line, the ridge
  emission has a flux of less than 1\% of the count rate of
  Swift\,1626.6$-$5156 during our observations (LO3). In the energy
  range around the first harmonic ($\sim$18\,keV, see
  \S\ref{sec:harmonic} and \S\ref{sec:phaseresolved}) no emission from
  the ridge is detected. Therefore Galactic ridge emission cannot
  explain the deviations that we measure.

\subsubsection{Alternative standard continuum and CRSF models}\label{sec:cont}

In order to test the continuum model dependence of our results we in
turn replaced the cutoff power law model by three other standard
empirical pulsar continuum models and repeated the fit for LO3. The
first of these was the power law with a high energy cutoff
(\texttt{power}$\times$\texttt{highecut} in \texttt{xspec} where, in
contrast to the smooth cutoff power law, the rollover in the spectrum
only starts at a certain energy, $E_\mathrm{cut}$ \citep{1983white}.
While this model is often used, it can create an artificial line-like
residual near $E_\mathrm{cut}$ due to the discontinuity at that energy
\citep[][and references therein]{2002coburn}. Next we fit the spectrum
with the Fermi-Dirac cutoff from \citet{1986tanaka}
(\texttt{power}$\times$\texttt{fdcut} in \texttt{xspec}). This model
has a smooth, continuous rollover described by $E_{\mathrm{cut}}$ and
$E_{\mathrm{fold}}$, which are, however, not directly comparable to
the values obtained with \texttt{cutoffpl} or
\texttt{power}$\times$\texttt{highecut} due to differences in the
continua. The final model we used is the Negative Positive Exponential
model from \citet{1995mihara} (\texttt{npex} in \texttt{xspec}), which
consists of two power laws with a smooth exponential cutoff described
by $E_{\mathrm{fold}}$. Each of these models was fit to LO3 modified
by absorption and with the added 6.4\,keV Gaussian emission line and
cyclotron line.  The fit values of the Gaussian emission line
  and the CRSF energy found for each model are given in
  Table~\ref{tab:new}.

These alternative continuum models were found to produce fits
  of similar quality when compared with each other and with the
  \texttt{cutoffpl} model. They also produced identical results,
  within errors, for all comparable parameters, most notably those of
  the 10\,keV CRSF. The $\Delta \chi^2$ test cannot be used to compare
  different continuum models \citep{protassov, 2012orlandini}. To
  quantify whether or not the \texttt{cutoffpl} model, which has the
  lowest $\chi^2_{\mathrm{red}}$ of the four continuum models tested,
  actually improves the continuum fit when compared to the alternative
  models,
we calculated the PCI (as in \S\ref{sec:best-fit}) of the
\texttt{cutoffpl} model. \citet{2012orlandini} discuss the
appropriateness of this $F$-test for comparison between different
models, and the statistic is applied in this way by
\citet{2012iwakiri}.  Table~\ref{tab:new} gives the reduced
$\chi^2_{\mathrm{red}}$, number of degrees of freedom, and PCI
(compared to \texttt{cutoffpl}) for each alternative model. We find
that the PCI is between $\sim$25--50\%, meaning that the
\texttt{cutoffpl} improves the continuum fit marginally at best.
Because none of the canonical models clearly gave a best fit to the
continuum, nor adjusted the energies of the 6.4\,keV line or the
10\,keV feature, and since \texttt{cutoffpl} has the smallest number
of parameters, we defaulted to this model, our original choice, as a
result of this test. The \texttt{cutoffpl} model is also well suited
for comparisons with many earlier results for similar sources
\citep{2002coburn, 2006mowlavi,2008caballero,2011suchy}.

Also, in order to further check that the $10\,$keV feature is robustly
described as a typical CRSF, we compared the Gaussian optical depth
absorption model with a Lorentzian profile model (\texttt{cyclabs} in
\texttt{xspec}) \citep{1990Mihara}. Taking the known systematic shift
into account we found similar centroid energies for the
absorption-like feature using either model. We chose to use
\texttt{gabs} in our analysis because its parametrization provides an
easier to interpret and compare characteristic energy parameter (in
contrast to \texttt{gabs} the energy parameter of \texttt{cyclabs}
does not directly reflect the energy where the absorption-like feature
in the spectrum is deepest). 

\subsubsection{Modified continuum models}\label{sec:modified}

Many accreting pulsar spectra are either not adequately or at
  least not uniquely described by the standard empirical continuum
  models above, especially in the 10--20\,keV range
  \citep{2002coburn,2004orlandini}. Incorrect modeling of the exact
  shape of the rollover can create residuals in addition to CRSF
  features or, in the worst case, mimic CRSFs. There are many examples
  for the application of continuum models with slightly modified
  cutoff characterization. \citet{1999orlandini} noted the likely
  presence of a two step change of the spectral slope for
  \objectname{OAO\,1657$-$415} (as well as a potential cyclotron line
  residual near 36\,keV), the first change occurring at 10--20\,keV
  and the other one at higher energies, however, they were able to
  model this behavior taking advantage of the discontinuity and the
  two characteristic energies of the \texttt{highecut} model mentioned
  above. Other authors used an additional polynomial component in
  order to describe a more complex rollover shape, i.e.,
  \citet{2000burderi} for \objectname{Cen~X-3} (an established
  cyclotron line source) and \citet{2008klochkov} for
  \objectname{EXO\,2030$+$375} (noting a potential cyclotron line
  residual near 63\,keV). The correct continuum and therefore the
  presence of a 25\,keV CRSF in \objectname{Vela X-1} (in addition to
  the established $\sim$55\,keV one) has long been debated
  \citep{1998orlandini,2002kreykenbohm,2003labarbera}. In the most
  recent analysis its presence was confirmed in a \textsl{Suzaku}
  observation \citep{2011doroshenko}. Another widely used modification
  is obtained by adding a broad Gaussian component, most often in
  emission, to a standard or already modified model
  \citep{2002coburn,2007klochkov,2008klochkov,2009ferrigno,2008suchy,2011suchy}.
  Since our Swift\,J1626.6$-$5156 observations show a positive
  residual at $\sim$14\,keV before modeling the cyclotron line
  (Figure~\ref{fig2}b) we applied this modified model to test whether
  our cyclotron line detection is robust against the choice of a more
  complex rollover description.

\begin{figure}
\includegraphics[width=\columnwidth]{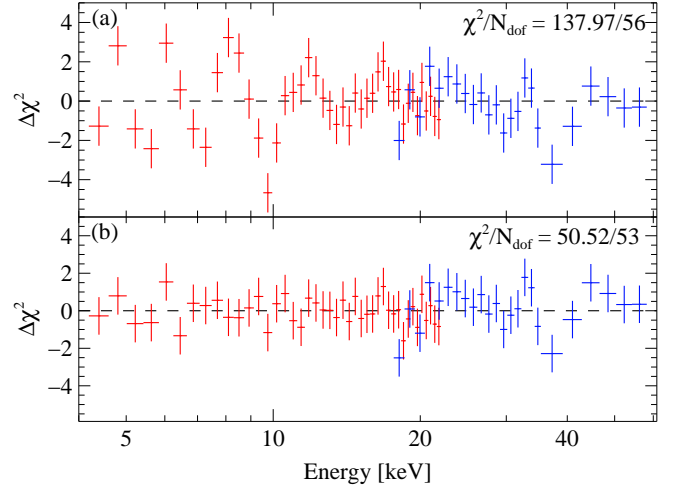}
\caption{\textit{a)} Starting from the continuum plus Fe line
    fit (Equation~\ref{eq1}) for observation LO3 (residuals shown in
    Figure~\ref{fig2}b), we further smoothed the continuum in this
    alternative fit by first including a broad Gaussian emission
    feature centered at 13.7\,keV. While this flattens the residuals
    substantially, the absorption-like feature at 10\,keV still
    remains. \textit{b)} Final fit for this mixed emission/absorption
    approach after also including the 9.7\,keV cyclotron
    line. The remaining residual near $\sim 36$\,keV is most likely 
    caused by an instrumental effect (\S\ref{sec:iodine}). \label{fig5}}
\end{figure}

We therefore altered our fit procedure by first adding a broad
emission line near $14\,$keV to the initial continuum plus Fe line fit
for LO3 (Figure~\ref{fig2}b, Equation~\ref{eq1}). The residuals of
this fit are shown in Figure~\ref{fig5}a. The negative residual near
10\,keV is still clearly visible, giving a $\chi^2_\mathrm{red}$ of
2.46 for 56 dof. Thus an absorption-like feature was still required
for an adequate description. Multiplying by a Gaussian optical depth
absorption feature and refitting we obtained $E_\mathrm{cyc} =
9.74^{+0.16}_{-0.12}$\,keV, $\sigma_{\mathrm{cyc}} =
0.93^{+0.22}_{-0.28}$\,keV, $\tau_{\mathrm{cyc}} =
0.14^{+0.04}_{-0.07}$ for the cyclotron line and $E_{\mathrm{em}} =
13.8^{+0.7}_{-1.6}$\,keV, $\sigma_{\mathrm{em}} =
1.86^{+0.98}_{-0.67}$\,keV, $A_{\mathrm{em}} = 0.26^{+0.33}_{-0.15}
10^{-3}$\,photons\,cm$^{-2}$\,s$^{-1}$ for the emission-like line.
This fit with both lines yielded $\chi^2/N_{\mathrm{dof}} = 49.96/53$
$(=0.94)$ (Figure~\ref{fig5}b), as opposed to $52.76/56$ $(=0.94)$ for
the original best-fit (Figure~\ref{fig2}c, Table~\ref{tab2}). The fits
are equally good (PCI\,$=$\,50\%). While the depth of the
cyclotron line was smaller in the modified fit, the line was still
detected at the same energy as before (Table~\ref{tab2}). The
parameters of the modified fit characterizing the continuum shape were
$\Gamma = 1.47^{+0.09}_{-0.11}$ and $E_{\mathrm{fold}} =
12.58^{+1.26}_{-1.32}$\,keV, consistent within errors with the
original fit (Table~\ref{tab2}). The same Fe line parameters were
obtained for both fit procedures.

This implies that, while the underlying continuum might be slightly
different from a standard cutoff power law, such a difference cannot
be significantly detected in this observation. The absorption-like
feature at 10\,keV on the other hand is significant. It appears at
9.7\,keV regardless of the presence or absence of a broad emission
feature and its energy is stable. Based on these results and since the
model without the broad emission feature is the simpler one, it is the
one we chose as our best-fit model. The same overall behavior has been
seen for all the datasets with the exception of DS3, for which a
better fit was obtained including a broad emission feature, see notes
of Table~\ref{tab2} for further details. This can be understood since
the DS3 spectrum was obtained averaging over many datasets, which
might contain some evolution of the continuum.

\subsubsection{18\,keV harmonic}\label{sec:harmonic}

Returning to Figure~\ref{fig2}c we point out a small depression near
18\,keV. In Figure~\ref{fig2}d we show the obtained residuals
after fitting this feature with another Gaussian optical depth
absorption line with centroid energy $18.5^{+0.9}_{-0.6}$\,keV, $\tau
\sim 0.1^{+0.09}_{-0.07}$, and $\sigma$, which could not be
constrained, fixed at $\sim$0.02. This additional component reduced
$\chi^2$ slightly but did not significantly improve the fit,
  as its PCI\,$\sim$\,40\% from the ratio of variances $F$-test
  \citep{2007press}.  We therefore did not include this feature in
our best-fit model of the phase averaged spectrum, i.e.,
Table~\ref{tab2}. While we would not normally consider this a
detection, we also found that the addition of this 18\,keV feature
greatly improved the pulse phase resolved spectral fits
(\S\ref{sec:phaseresolved}) during the pulse peak. It is
marginally present in the phase averaged spectrum of LO2 as
well. It is thus likely the first harmonic of the fundamental CRSF, which is significantly detected in only a few profile phase bins (see \S\ref{sec:phaseresolved}).

\subsubsection{36\,keV instrumental feature}\label{sec:iodine}

Even after fitting the harmonic feature there remained an even more
obvious line-like residual near 36\,keV (Figures~\ref{fig2}d and
\ref{fig5}b). It is tempting to interpret this as a further
harmonic. There might even be another weak residual at
$\lesssim$30\,keV. However, the effective area of HEXTE shows a sharp
drop at 33\,keV due to the K-edge of iodine with a smooth recovery
until about 45\,keV\citep[see Figure~5 of][]{1998rothschild}. So there
is a good chance that (at least part of) any residuals in this energy
range (is) are due to imperfect calibration \citep{1999heindl_a}. In
addition the strongest internal HEXTE background line, which is due to
K-lines from the tellurium daughters of various iodine decays, sits at
30\,keV \citep{1998rothschild}. We therefore decided to not further
pursue the potential existence of harmonics beyond the first one at
18\,keV. If Swift\,J1626.5$-$5156 shows other outbursts in the future
the detection of these harmonics might be possible with higher
sensitivity with \textsl{Suzaku}, or better, \textsl{NuSTAR} or
\textsl{Astro-H}.

\begin{figure}
\includegraphics[width=\columnwidth]{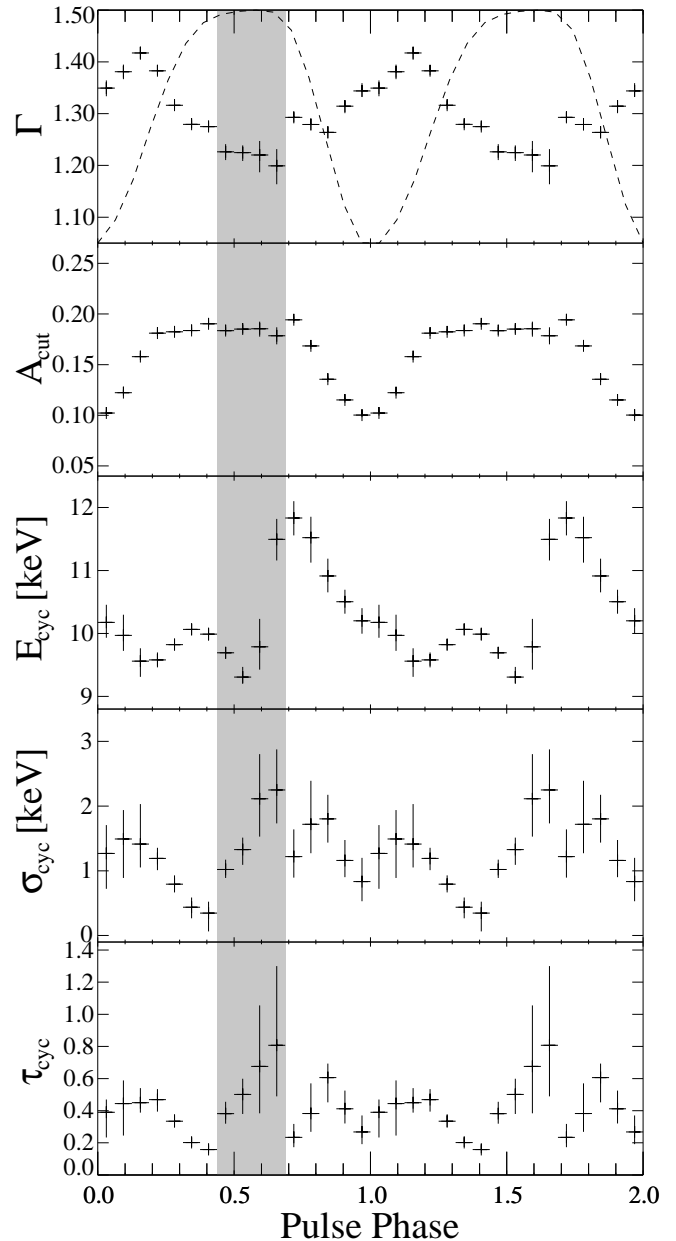}
\caption{Variability of the spectral parameters with phase for
  observation LO3. The top panel includes the 3.5--25\,keV pulse
  profile of Swift\,J1626.6$-$5156 (dashed line) as well as the
  best-fit power law index (crosses). The shaded region indicates the
  four phase bins surrounding the pulse peak, in which a second CRSF
  line is necessary to produce an acceptable fit. \label{fig6}}
\end{figure}

\subsection{Pulse phase resolved spectra}\label{sec:phaseresolved}

We again used LO3, which shows the strongest cyclotron line, for a
phase resolved spectral analysis. We extracted Good Xenon event data
with a time resolution of 3.9\,ms and corrected the event times to the
Solar System barycenter. A period search on these data resulted in a
pulse period of $15.35828\pm0.00008$\,s, consistent with the spin down
evolution presented by \citet{2010baykal}. We folded the 3.5--25\,keV
events on this period in order to obtain the pulse profile, see dashed
line in the upper panel of Figure~\ref{fig6}. As presented by
\citet{2008reig} the pulse profile shows a single broad peak with a
roughly constant maximum over four phase bins. The falling edge seems
to be steeper than the rising edge, which may be an artifact from the
period folding. We did not include a binary correction since effects
of the $\sim$133\,d orbit \citep{2010baykal} are negligible for our
$\lesssim$30\,ks observation.

We divided the pulse profile into 16 phase bins for spectral analysis
and used \texttt{ikfasebin} to extract spectra for each
phase\footnote{See
  http://pulsar.sternwarte.uni-erlangen.de/wilms/research/analysis/
  rxte/pulse.html for more detailed information.}. Each spectrum was
fit using Eq.~\ref{eq2} with an additional edge component at
$\sim$5\,keV to account for the Xenon L-edge calibration feature
\citep{2006jahoda}. Still we found it necessary to remove one spectral
bin at $\sim$4.5\,keV (spectral bin 12) from each phase's spectrum to
improve the overall fit. We assume that the systematic residual in
this spectral bin is due to calibration uncertainties at lower
energies and has no physical significance for the model. The column
density and the iron emission energy were frozen at $3.4 \times
10^{22}\,\mathrm{cm}^{-2}$ and 6.4\,keV -- the values from the phase
averaged fit --, respectively. In order to better constrain the
spectral index, the folding energy of the cutoff power law was also
frozen at the phase averaged value of 10.98\,keV.

\begin{deluxetable*}{ccccccccccccccc}
\tablecaption{Spectral parameters for phase resolved spectra from LO3
  \label{tab3}}
\tabletypesize{\scriptsize}
\scriptsize
\tablewidth{0pt}
\tablecolumns{15} \tablehead{
\colhead{Phase} & \colhead{$\Gamma$} & \colhead{$A_\mathrm{cut}$} &
\colhead{$\sigma_\mathrm{Fe}$} & \colhead{$A_\mathrm{Fe}$} &
\colhead{$E_\mathrm{cyc,1}$} & \colhead{$\sigma_\mathrm{cyc,1}$} & \colhead{$\tau_\mathrm{cyc,1}$} & \colhead{$E_\mathrm{cyc,2}$} & \colhead{$\sigma_\mathrm{cyc,2}$} & \colhead{$\tau_\mathrm{cyc,2}$} & \colhead{$\chi^2_\mathrm{red,2}$} & \colhead{$\chi^2_\mathrm{red,1}$} & \colhead{$\chi^2_\mathrm{red,0}$} & \colhead{PCI$^c$}\\
\colhead{Bin} & \colhead{} & \colhead{[$10^{-1}$]$^a$} & \colhead{[keV]} & \colhead{[$ 10^{-3}$]$^b$} & \colhead{[keV]} & \colhead{[keV]} & \colhead{} & \colhead{[keV]} & \colhead{[keV]} & \colhead{} & \colhead{41 dof } & \colhead{43 dof } & \colhead{46 dof } & \colhead{}
} \startdata

$1$  & $1.34^{+0.02}_{-0.01}$  & $1.02^{+0.03}_{-0.03}$  &
$0.3^{+0.1}_{-0.3}$  & $1.02^{+0.50}_{-0.34}$  
     & $10.2^{+0.3}_{-0.2}$    & $1.3^{+0.4}_{-0.5}$  & $0.4^{+0.2}_{-0.1}$  & $--$  & $--$  & $--$  
     & $--$  &\textbf{1.34}  & $2.76$ & $9.2 \times 10^{-3}$ \\[1mm]   

$2$  & $1.38^{+0.02}_{-0.01}$  & $1.22^{+0.03}_{-0.03}$  &
     $0.3^{+0.2}_{-0.2}$  & $1.15^{+0.75}_{-0.19}$  
     & $10.0^{+0.3}_{-0.2}$  & $1.6^{+0.5}_{-0.6}$  & $0.5^{+0.1}_{-0.2}$  & $--$  & $--$  & $--$  
     & $--$  &\textbf{0.98}  & $2.10$ & $6.5 \times 10^{-3}$ \\[1mm]   

$3$  & $1.42^{+0.01}_{-0.01}$  & $1.57^{+0.03}_{-0.03}$  &
     $0.3^{+0.2}_{-0.3}$  & $1.32^{+0.69}_{-0.38}$  
     & $9.6^{+0.2}_{-0.1}$  & $1.4^{+0.6}_{-0.4}$  & $0.5^{+0.2}_{-0.1}$  & $--$  & $--$  & $--$  
     & $--$  &\textbf{0.93}  & $2.41$ & $1.0 \times 10^{-3}$ \\[1mm]   

$4$  & $1.38^{+0.01}_{-0.01}$  & $1.81^{+0.04}_{-0.03}$  &
     $0.0^{+0.0}_{-0.0}$  & $1.10^{+0.29}_{-0.20}$  
     & $9.6^{+0.1}_{-0.1}$  & $1.2^{+0.2}_{-0.2}$  & $0.5^{+0.0}_{-0.1}$  & $--$  & $--$  & $--$  
     & $--$  &\textbf{0.68}  & $4.64$ & $1.8 \times 10^{-9}$ \\[1mm]   

$5$  & $1.32^{+0.01}_{-0.01}$  & $1.82^{+0.03}_{-0.03}$  &
     $0.0^{+0.01}_{-0.00}$  & $1.15^{+0.26}_{-0.24}$  
     & $9.8^{+0.1}_{-0.1}$  & $0.8^{+0.1}_{-0.1}$  & $0.3^{+0.0}_{-0.0}$  & $--$  & $--$  & $--$  
     & $--$  &\textbf{0.96}  & $6.57$ & $1.7 \times 10^{-9}$ \\[1mm]   

$6$  & $1.28^{+0.01}_{-0.01}$  & $1.83^{+0.04}_{-0.03}$  &
     $0.8^{+0.1}_{-0.1}$  & $2.75^{+ 0.52}_{-0.47}$  
     & $10.0^{+0.2}_{-0.0}$  & $0.4^{+0.2}_{-0.1}$  & $0.2^{+0.0}_{-0.0}$  & $--$  & $--$  & $--$  
     & $--$  &\textbf{1.69}  & $5.46$ & $8.6 \times 10^{-9}$ \\[1mm]   

$7$  & $1.28^{+0.00}_{-0.01}$  & $1.90^{+0.03}_{-0.03}$  &
     $1.0^{+0.1}_{-0.2}$  & $2.96^{+0.51}_{-0.46}$  
     & $10.0^{+0.1}_{-0.1}$  & $0.3^{+0.2}_{-0.3}$  & $0.2^{+0.0}_{-0.1}$  & $--$  & $--$  & $--$  
     & $--$  &\textbf{1.41}  & $4.35$ & $1.5 \times 10^{-4}$ \\[1mm]   

$8$  & $1.23^{+0.01}_{-0.02}$  & $1.84^{+0.03}_{-0.04}$  &
     $0.0^{+0.4}_{-0.0}$  & $0.62^{+0.26}_{-0.19}$  
     & $9.7^{+0.1}_{-0.1}$  & $1.0^{+0.2}_{-0.1}$  & $0.4^{+0.1}_{-0.6}$  & $17.5^{+0.6}_{-0.4}$  
     & $2.5$  & $0.8^{+0.2}_{-0.2}$  &\textbf{1.24} & $2.29$  & $5.38$ & $2.5 \times 10^{-2}$ \\[1mm]   

$9$  & $1.22^{+0.02}_{-0.01}$  & $1.85^{+0.04}_{-0.04}$  &
     $0.0^{+0.3}_{-0.0}$  & $0.74^{+0.49}_{-0.20}$  
     & $9.3^{+0.2}_{-0.1}$  & $1.3^{+0.2}_{-0.2}$  & $0.5^{+1.0}_{-0.1}$  & $18.4^{+0.4}_{-0.4}$  
     & $2.5$  & $1.0^{+0.3}_{-0.2}$  &\textbf{1.05} & $2.80$  & $5.42$ & $1.0 \times 10^{-3}$ \\[1mm]   

$10$ & $1.22^{+0.03}_{-0.3}$  & $1.85^{+0.07}_{-0.08}$  &
     $0.4^{+0.1}_{-0.1}$  & $2.28^{+1.14}_{-0.39}$  
     & $9.8^{+0.4}_{-0.4}$  & $2.1^{+0.7}_{-0.6}$  & $0.7^{+0.4}_{-0.3}$  & $18.7^{+0.3}_{-0.4}$  
     & $2.5$  & $1.3^{+0.5}_{-0.3}$  &\textbf{1.33}  & $2.81$  & $3.64$ & $8.8 \times 10^{-3}$ \\[1mm]   

$11$ & $1.20^{+0.03}_{-0.03}$  & $1.78^{+0.08}_{-0.08}$  &
     $0.4^{+0.1}_{-0.0}$  & $4.32^{+1.16}_{-1.02}$  
     & $11.5^{+0.3}_{-0.3}$  & $2.2^{+0.6}_{-0.5}$  & $0.8^{+0.5}_{-0.3}$  & $19.1^{+0.4}_{-0.5}$  
     & $2.5$  & $1.5^{+0.4}_{-0.5}$  &\textbf{1.17}  & $2.17$  & $2.62$ & $2.5 \times 10^{-2}$ \\[1mm]   

$12$ & $1.29^{+0.01}_{-0.01}$  & $1.94^{+0.03}_{-0.03}$  &
     $0.4^{+0.1}_{-0.0}$  & $4.40^{+0.68}_{-0.71}$  
     & $11.8^{+0.3}_{-0.3}$  & $1.2^{+0.4}_{-0.3}$  & $0.2^{+0.1}_{-0.0}$  & $--$  & $--$  & $--$  
     & $--$  &\textbf{1.43}  & $2.58$ & $2.7 \times 10^{-2}$ \\[1mm]   

$13$ & $1.28^{+0.01}_{-0.01}$  & $1.68^{+0.03}_{-0.03}$  &
     $0.4^{+0.1}_{-0.1}$  & $2.44^{+0.85}_{-0.69}$  
     & $11.5^{+0.3}_{-0.4}$  & $1.7^{+0.7}_{-0.4}$  & $0.4^{+0.2}_{-0.1}$  & $--$  & $--$  & $--$  
     & $--$  &\textbf{0.89}  & $2.25$ & $1.3 \times 10^{-3}$ \\[1mm]   

$14$ & $1.26^{+0.01}_{-0.01}$  & $1.35^{+0.03}_{-0.03}$  &
     $0.3^{+0.1}_{-0.3}$  & $1.13^{+0.71}_{-0.34}$  
     & $10.9^{+0.3}_{-0.3}$  & $1.8^{+0.4}_{-0.4}$  & $0.6^{+0.2}_{-0.1}$  & $--$  & $--$  & $--$  
     & $--$  &\textbf{1.03}  & $3.66$ & $2.6 \times 10^{-5}$ \\[1mm]   

$15$ & $1.31^{+0.01}_{-0.01}$  & $1.15^{+0.02}_{-0.03}$  &
     $0.3^{+0.2}_{-0.2}$  & $1.18^{+0.45}_{-0.27}$  
     & $10.5^{+0.2}_{-0.2}$  & $1.2^{+0.3}_{-0.3}$  & $0.4^{+0.1}_{-0.1}$  & $--$  & $--$  & $--$  
     & $--$  & \textbf{1.02}  & $3.49$ & $4.2 \times 10^{-5}$ \\[1mm]   

$16$ & $1.34^{+0.02}_{-0.01}$  & $1.00^{+0.03}_{-0.02}$  &
     $0.3^{+0.2}_{-0.1}$  & $1.31^{+0.46}_{-0.34}$  
     & $10.2^{+0.2}_{-0.2}$  & $0.8^{+0.4}_{-0.3}$  & $0.3^{+0.1}_{-0.1}$  & $--$  & $--$  & $--$  
     & $--$  &\textbf{0.99}  & $2.38$ & $2.2 \times 10^{-3}$ \\[1mm]   
     
\enddata \tablecomments{$A_\mathrm{cut}$ and $A_\mathrm{Fe}$ are
  respectively the normalization constants of the cutoff power law and
  Gaussian iron emission line; $E_{\mathrm{cyc,1}}$,
  $\sigma_{\mathrm{cyc,1}}$, and $\tau_{\mathrm{cyc,1}}$ the energy,
  width, and Gaussian optical depth of the first CRSF;
  $E_{\mathrm{cyc,2}}$, $\sigma_{\mathrm{cyc,2}}$, and
  $\tau_{\mathrm{cyc,2}}$ the same parameters for the second CRSF;
  $\chi_{\mathrm{red,0}}$, $\chi_{\mathrm{red,1}}$, and
  $\chi_{\mathrm{red,2}}$ the reduced $\chi^2$ values for fits with no
  CRSF, one (the fundamental) CRSF, and both CRSFs. $^a$Units are
  $10^{-1}$ photons\,keV$^{-1}$\,cm$^{-2}$ at $1$\,keV.  $^b$Units are
  $10^{-3}$ total photons\,cm$^{-2}$\,s$^{-1}$ in the line.
  $N_\mathrm{H}$, $E_\mathrm{Fe}$, and $E_\mathrm{fold}$ were frozen
  to their phase averaged values. $^c$As stated in the text in
    \S\ref{sec:phaseresolved}, the probability of chance improvement
    (PCI) is calculated between the models with no CRSF and one CRSF
    for phase bins 1--7 and 12--16, and between one CRSF and two CRSFs
    for phase bins 8--11 (the latter are listed in bold print). The PCI
    values between models with no CRSF and one CRSF for bins 8, 9, 10,
    and 11 are respectively $2.8 \times 10^{-3}$, $1.5 \times
    10^{-2}$, $0.20$, and $0.27$. The PCI is described briefly in
    \S\ref{sec:best-fit} and \S\ref{sec:cont}.  }
  
\end{deluxetable*}

Using only the continuum plus Fe line model (Eq.~\ref{eq1}) on the
phase resolved spectra led to unacceptable fits throughout the whole
pulse, see column $\chi^2_{\mathrm{red,0}}$ in Table~\ref{tab3}.
Including a CRSF in the form of a Gaussian optical depth absorption
line at $\approx 10$\,keV (Eq.~\ref{eq2}) improved the fits
significantly, but did not result in a sufficiently low $\chi^2$
during the maximum of the main peak, see column
$\chi^2_{\mathrm{red,1}}$ in Table~\ref{tab3}. Including a second
absorption line in these four phase bins improved the fits to
acceptable values, see column $\chi^2_{\mathrm{red,2}}$ in
Table~\ref{tab3}. The results from these fits are shown in
Figure~\ref{fig6} and Table~\ref{tab3}. The PCI
  \citep{2007press} in the last column was calculated for the best fit
  in each phase bin, i.e., between the models with no CRSF and one
  CRSF for phase bins 1--7 and 12--16, and between no CRSF and two
  CRSFs for phase bins 8--11. The four phase bins requiring the
second cyclotron line are highlighted by the shaded region in
Figure~\ref{fig6}. The width of the second line was frozen at
$2.5$\,keV in order to better constrain the other parameters. The
centroid energy is roughly twice the energy of the fundamental line
and is consistent with being its first harmonic. The phase
  resolved results for the harmonic likely explain the additional
  feature mentioned in the description of the phase averaged spectrum
  (\S\ref{sec:harmonic}, Figure~\ref{fig2}d). It has such a low
  significance in the phase averaged spectrum because it is not
  detected at most rotation phases and its energy changes with
  phase. There are other CRSFs that are only detected in phase
  resolved spectra.  For example, the second harmonic of the CRSF in
  Her X-1 is detected only in the descent of the main peak
  \citep{2004diSalvo, 2008enoto}.

Figure~\ref{fig6} shows that the power law index hardens throughout
the peak and softens in the minimum. As can be expected the power law
normalization roughly follows the pulse profile, reaching
$\sim$$0.18\,\mathrm{photons}\,\mathrm{cm}^{-2}\,\mathrm{keV}^{-1}$ in
the peak. The CRSF energy shows a smaller increase during the rising
edge, then first a decline and then a drastic increase of more than
20\% in the peak itself, followed by a slow decline over the
  falling edge and minimum of the pulse profile. The width and depth
  of the CRSF are variable as well with a dip in the rising edge and a
  possible increase through the peak and a rapid decline at the
  beginning of the falling edge.

\section{Discussion}\label{sec:discussion}

\subsection{Magnetic field estimated from $E_\mathrm{cyc}$}\label{sec:bfield1}

We found evidence for cyclotron resonance scattering features at
$\sim$10 and 18\,keV in the spectrum of the Be/X-ray binary
Swift\,J1626.6$-$5156. While the harmonic CRSF was not seen at a
significant level in the phase averaged spectrum, pulse phase resolved
spectroscopy showed it is a necessary addition to the spectral fit
between phases $\sim$0.4--0.7, i.e., during the pulse peak. We can use
the cyclotron line energy to estimate the strength of the magnetic
field local to the line production region near the neutron star's
polar regions. The fundamental and subsequent harmonic cyclotron line
energies are related to the local magnetic field strength by
\citep{1992meszaros}
\begin{eqnarray}
E_n & = m_{\mathrm{e}} c^2 \frac{\sqrt{1+2n(B/B_{\mathrm{crit}}) \sin^2 \theta} - 1}{\sin^2 \theta} \frac{1}{1+z} \\
\nonumber & \approx n (11.6\,{\mathrm{keV}})(1+z)^{-1} B_{12}
\end{eqnarray}
where $m_{\mathrm{e}}$ is the electron rest mass, $n = (1, 2, 3,
\ldots)$ is the integer harmonic number, $B_{\mathrm{crit}} \sim 4
\times 10^{13}$\,G is the critical field strength for resonant
scattering, $\theta$ is the angle between the photon direction and the
magnetic field vector, $z$ is the gravitational redshift at the neutron
star surface ($z \sim 0.3$ for a neutron star with mass
$1.4\,M_{\odot}$ and radius $\sim 10^6$\,cm) and $B_{12} = B
/(10^{12}\,G)$. A fundamental energy of $\sim$10\,keV therefore
implies a magnetic field of $\sim 8.6 (1+z) 10^{11}$\,G $\sim 1.1
\times 10^{12}$\,G.

\subsection{Magnetic field estimated from accretion}\label{sec:bfield2}

As a consistency check we estimated the global magnetic field
strength, assuming that the neutron star has reached its equilibrium
spin rate through interplay between an accretion disk and the magnetic
field. We refer to \citet{ghosh2} and \citet[][Chapter~15]{shapiro}
for our calculation and use the pulse period of LO3. While the pulse
period evolved over the outburst, the change was small enough to not
throw the spin out of equilibrium: we refer to Figure 3 of
\citet{2010baykal}, which shows a change of only 0.2\% in the pulsar's
spin period over the whole activity phase. This small change in
$P_{\mathrm{spin}}$ had little effect on our estimate of the global
magnetic field. We estimated $B_{\mathrm{global}}$ by equating the
Alfv\'{e}n radius $r_{\mathrm{A}}$ with the corotation radius at which
the orbital velocity is equivalent to the surface velocity of the
rotating star, thus giving $\omega_{\mathrm{spin}} \approx \sqrt{GM /
  r_{\mathrm{A}}^3}$\,rad\,s$^{-1}$ where
\begin{equation}
r_{\mathrm{A}} \approx 3.2 \times 10^8 \dot{M}_{17}^{-2/7} \mu_{30}^{4/7}
\left ( \frac{M}{M_{\odot}} \right ) ^{-1/7} \, \mathrm{cm}
\label{rA}
\end{equation}
Here $\dot{M}_{17} = \dot{M}/(10^{17}\,\mathrm{g}\,\mathrm{s}^{-1})$
is the mass accretion rate and $\mu_{30} = \mu /
(10^{30}\,\mathrm{G}\,\mathrm{cm}^3)$ is the dipolar magnetic moment.
$P_{\mathrm{spin}} = 15.35828$\,s is our measured spin period, so
$\omega_{\mathrm{spin}} \sim 0.41\,\mathrm{rad}\,\mathrm{s}^{-1}$ and
$r_{\mathrm{A}} \sim 1 \times 10^9$\,cm. To estimate the average
$\dot{M}$ over the source's lifetime, we must first estimate the
luminosity of the source outside of outburst. \citet{2011reig}
estimated a source distance of $d = 10.7 \pm 3.5$\,kpc. We assume that
the current low luminosity state of the source is typical outside of
outburst and from the source's spectrum during this state we derived a
2--60\,keV flux of $2.9 \times
10^{-11}\,\mathrm{erg}\,\mathrm{cm}^{-2}\,\mathrm{s}^{-1}$, giving a
luminosity of $\sim 1.7 \times 10^{35}\,\mathrm{erg}\,\mathrm{s}^{-1}$
at 10\,kpc for isotropic emission. The mass accretion rate for an
efficiency $\eta \approx 0.1$ is then $\dot{M} = L/(\eta c^2) \sim 2
\times 10^{15}\,\mathrm{g}\,\mathrm{s}^{-1}$. Using Equation~\ref{rA}
we find $\mu_{30} \sim 1$ and the global dipolar magnetic field
strength $B_\mathrm{global} = 2 \mu / R^3_\mathrm{NS} = 2 \times
10^{12}$\,G. As a rough estimate this is reasonably close to the field
strength derived from the cyclotron line (see previous section).

\subsection{$E_\mathrm{cyc}$ dependence on luminosity}\label{sec:luminosity}

Something else to consider is the relationship between the cyclotron
line energy and the source's X-ray luminosity. \objectname{Her~X-1}
and \objectname{GX~304$-$1} are examples of systems containing an
accreting magnetized neutron star whose cyclotron line energy is
positively correlated with luminosity
\citep{2007staubert,2012klochkov}. A counter example is
\objectname{V\,0332$+$53}, in which the cyclotron line energy
increases as the source's luminosity decays and vice versa
\citep{2006mowlavi,2010tsygankov}\footnote{\objectname{4U\,0115$+$63}
  is currently also considered to be an example for a negative
  $E_\mathrm{cyc}$-$L$ correlation \citep{2006nakajima}. In this case
  the presence of the correlation, however, depends on the continuum
  model \citep{2011mueller}.}. It was suggested by
\citet{2007staubert} that a positive correlation between
$E_\mathrm{cyc}$ and luminosity would occur for systems accreting at a
sub-Eddington rate, while the negative correlation would occur when
the accretion is locally super-Eddington. They show that the
fractional change in $E_\mathrm{cyc}$ is directly proportional to the
fractional change in $L$. If such a relationship could be confirmed
with cyclotron line sources of known distances it would be possible to
use observations of cyclotron lines as standard candles.

The 90\% errors on the cyclotron line energy in Table~\ref{tab2} point
to a possible correlation between $E_\mathrm{cyc}$ and luminosity. We
calculated the $3 \sigma$ errors on the line energy of LO1, LO3, and
DS3, observations which are clearly separated in time, and found that
the energies are not consistent with each other within the errors. The
$3 \sigma$ confidence range on the CRSF energy is 10.15--10.55\,keV
during LO1, 9.58--9.85\,keV during LO3, and 9.32--9.56\,keV during
DS3. The measured decrease in the line energy with decreasing flux
(luminosity) is indicative of a positive correlation between
$E_\mathrm{cyc}$ and $L$ and is apparent at the 99.7\% confidence
level in our data. Further observations during a future outburst are
necessary to confirm this correlation.

\citet{2012becker} recently derived a new expression for the critical
luminosity assuming a simple physical model for the accretion column.
According to their estimate, and using the $B$ field determined in
\S\ref{sec:bfield1}, the critical luminosity for Swift\,J1626.5$-$5156
is $L_\mathrm{crit} = 1.5 \times 10^{37}
B_{12}^{16/15}\mathrm{erg}\,\mathrm{s}^{-1}$ $= 1.7 \times
10^{37}\mathrm{erg}\,\mathrm{s}^{-1}$. The normalized 3--20\,keV
luminosity for LO1, the observation with the highest flux $F$
(Table~\ref{tab2}), is $L/L_\mathrm{crit} = 4\pi d^2F /
L_\mathrm{crit} = 1.6$ at a distance $d=10.7$\,kpc, with all
observations from DS1 and later falling below $L_\mathrm{crit}$. Thus
Swift\,J1626.6$-$5156 was mainly observed in the sub-critical
accretion regime and the observed positive $E_\mathrm{cyc}$-$L$
correlation is consistent with the theoretical expectation.

\subsection{CRSF parameter dependence on pulse phase}

The strong $\gtrsim$20\% variation of the fundamental energy with
pulse phase that we measured for Swift\,J1626.6$-$5156 is consistent
with several other sources, namely Cen~X-3
\citep{2000burderi,2008suchy}, Vela~X-1
\citep{2003labarbera,2002kreykenbohm}, 4U\,0115$+$63
\citep{2004heindl}, and GX~301$-$2 \citep{2004kreykenbohm,2012suchy}.
In all cases the fundamental energy varies by about 10--30\% over the
pulse. This variability has been attributed to the change in
  viewing angle and accretion column throughout the pulse, resulting
  in a variable local magnetic field as a function of
  pulse phase. Specifically the $E_\mathrm{cyc}$ versus phase profile
  of Cen~X-3 has very similar properties to the one of
  Swift\,J1626.6$-$5156 (though shifted in phase from the falling to
  the rising edge of the pulse): a sharp rise followed by a slower
  decay. For Cen~X-3 these $E_\mathrm{cyc}$ variations were shown to
  be inconsistent with a changing viewing angle on a pure dipole
  magnetic field but it was found that it is possible that emission
  from above both poles is observed \citep{2008suchy}. The width and
  depth of the line are also generally variable over pulse phase,
  similar to what was observed for Swift\,J1626.6$-$5156. We interpret
  the phase dependence of the CRSF parameters as additional evidence
  that the feature is real and not an artifact of the spectral
  fitting.

\subsection{Line spacing}\label{sec:spacing}

The ratio between the fundamental and harmonic line energies for the
phase averaged fit is 1.9 and ranges from 1.7 to 2.0 for the four
phase bins where the harmonic line is detected. Thus the harmonic probably 
does not occur at strictly twice the fundamental energy. Other
sources, for example V\,0332$+$53
\citep{2005pottschmidt,2005kreykenbohm,2010nakajima} and 4U\,0115$+$63
\citep{1999heindl_b,1999santangelo,2006nakajima}, have also displayed
cyclotron lines at non-integer multiples of the fundamental
energy. Spacing that is narrower than in the harmonic case is expected
when general relativistic effects are taken into account
\citep{1992meszaros}, while a wider spacing has, e.g., been explained
as being due the emission regions of the two lines being located at
different heights in the accretion column \citep{2010nakajima}. For
Swift\,J1626.6$-$5156 the relativistic effect might have been
observed, similar to V\,0332$+$53
\citep{2005pottschmidt,2005kreykenbohm}\footnote{For V\,0332$+$53 the
  situation is not clear, however, since ratios $>$2 have also been obtained
  \citep{2010nakajima}, depending on the line and continuum model .}.

\subsection{Iron line}\label{sec:iron}

Swift\,J1626.6$-$5156 is located in the direction of the Galactic
plane, so the diffuse Galactic Fe K emission must at least contribute
to our measured iron line flux. From the quiescence observations we
already know that this contribution is small (\S\ref{sec:ridge}). In
addition we can make the following estimate: For a typical Galactic
ridge location, the total Fe K line flux is $\sim 3.8 \times
10^{-4}\,\mathrm{photons}\,\mathrm{s}^{-1}\,\mathrm{cm}^{-2}\,\mathrm{deg}^{-2}$
\citep{2008ebisawa}. The PCA solid opening angle is 0.975\,deg$^2$ and
the effective area of one PCU at 6.4\,keV is $1000\,\mathrm{cm}^2$, so
the diffuse Fe K line flux (averaged over the area) seen by the PCA is
$\sim$$0.37\,\mathrm{photons}\,\mathrm{s}^{-1}$ \citep{2006jahoda}. We
calculated the 3--20\,keV Fe K line flux from each of the three long
observations by first finding the flux of the best-fit model and then
removing the emission line and, without refitting, calculating the
flux again. We find Fe K fluxes of
$\sim$$3.0\,\mathrm{photons}\,\mathrm{s}^{-1}$ from LO1,
$\sim$$2.7\,\mathrm{photons}\,\mathrm{s}^{-1}$ from LO2, and
$\sim$$1.0\,\mathrm{photons}\,\mathrm{s}^{-1}$ from LO3, indicating
that most of the iron line emission is coming from
Swift\,J1626.6$-$5156. We further note that the flux of the iron line,
$A_\mathrm{Fe}$, as well as of the absorption $N_\mathrm{H}$ decrease
over the time of the outburst (Table~\ref{tab2}). Since $L \propto
\dot{M}$ and since $\dot{M}$ depends on the amount of material in the
system, this may reflect the diminishment of material from the neutron
star's surroundings.

\section{Conclusion}\label{sec:conclusion}

We have clearly detected a spectral feature at $\sim$10\,keV in the
spectrum of the Be/X-ray binary Swift\,J1626.6$-$5156. Every aspect of
our analysis points to this feature being a cyclotron resonance
scattering feature. The parameters of the cyclotron line vary
  with pulse phase in a manner consistent with other accreting pulsars
  displaying cyclotron lines. The second harmonic of the CRSF, at an
energy of $\sim$18\,keV, is seen between pulse phases 0.4--0.7. The
fundamental cyclotron line and the $\sim$6.4\,keV iron emission line
persist throughout the source's outburst and oscillatory stages.

The neutron star's magnetic field strength of $\sim$$10^{12}$\,G,
derived from the cyclotron line energy, is physically realistic and
in the expected range for high-mass X-ray binaries. Interestingly 
it is among the lowest $B$ fields measured for an accreting pulsar 
so far,  together with that of 4U\,0115$+$63 \citep{2006nakajima}.
We find that this field strength is roughly consistent with the scenario 
of spin equilibration via disk accretion and we do not rule out the 
possibility that Swift\,J1626.6$-$5156 may host an accretion disk.

Comparing the $3\sigma$ confidence intervals of the fundamental CRSF
energies in LO1, LO3, and DS3 reveals a positive correlation between
the line energy and the source luminosity. This behavior is similar to
that of Her X-1 and GX~304$-$1, in which the luminosity and line
energy are also positively correlated, and according to the model of
\citet{2007staubert} and \citet{2012becker} implies that the outburst
decay of Swift\,J1626.6$-$5156 was sub-Eddington. Observations during
future outbursts of Swift\,J1626.6$-$5156 will be beneficial in
confirming this relationship between its luminosity and cyclotron line
energy.

\begin{acknowledgments}
We thank W.\ Coburn for pointing out Swift\,J1626.6$-$5156 as a
candidate cyclotron line source. We acknowledge the useful
  comments of the referee that substantially improved this paper, and
  thank him or her for pointing out the use of the ratio of variances
  $F$-test from which we derived the probability of chance improvement
  for a given model. This work was funded under solicitation
NNH07ZDA001N-SWIFT4 of the Swift Guest Investigator program and NSF
Grant AST0708424. J.\ Wilms acknowledges partial funding from the
Bundesministerium f\"ur Wirtschaft und Technologie through Deutsches
Zentrum f\"ur Luft- und Raumfahrt grants 50 OR 0808 and 50 OR
0905. S.\ Suchy acknowledges the support of NASA contract NAS5-30720
and NASA grant NNX08AZ82G.
\end{acknowledgments}

\end{document}